\documentclass[a4paper]{jpconf}
\usepackage{graphicx}
\begin{document}
\title{Aspects of the T-duality construction for the Supermembrane theory.}

\author{M.P. Garcia del Moral$^1$, J.M. Pena$^2$, A. Restuccia$^1$}

\address{$^1$Departamento de F\'{\i}sica, Universidad de Antofagasta, Aptdo 02800, Chile.\\
  $^2$ Departamento de F\'{\i}sica,  Facultad de Ciencias,\\
 Universidad Central de Venezuela, A.P. 47270, Caracas 1041-A, Venezuela}

\ead{maria.garciadelmoral@uantof.cl; jpena@fisica.ciens.ucv.ve;alvaro.restuccia@uantof.cl}

\begin{abstract}
In this note we explicitly show how the generalization of the T-duality symmetry of the supermembrane theory compactified in $M_9\times T^2$ can be reduced to a parabolic subgroup of $SL(2,Z)$ that acts non-linearly on the moduli parameters and on the KK and winding charges of the supermembrane.  This is a first step towards a deeper understanding of the dual relation between the parabolic type II gauged supergravity in nine dimensions.
\end{abstract}

\section{Introduction}
There is a renewed interest in finding a T-duality invariant target space formulation of String theory \cite{musaev, aldazabal, nunez} and respectively a U-duality invariant action of M-theory \cite{6hullreid,bt,bmt,daniel,bmp}. Indeed, since String Theories and their corresponding effective Supergravity actions are limits of M-theory connected through dualities, M-theory should contain dualities as symmetries of the theory. This idea has been mainly realized in a bottom-up approach, by generalizing the geometry through the introduction of extra coordinates associated to the winding modes in supergravity actions in the context of Double Field Theory \cite{hz,reids, stw}. This approach was originally introduced by \cite{duff, tseytlin, seigel}. The reduction to normal coordinate space leads to gauged supergravity actions \cite{24hull-tfolds}.
 In this note we follow a different approach to realize it: we deep on the study of the U-duality invariant action of the Supermembrane worldvolume action \cite{gmpr2,tduality,gm}. This search is an intermediate step, to obtain a generalization of T-duality for M-theory, in particular for supermembrane theory \cite{tduality, gm}. This generalization acts locally on the theory and in the string theory limit one recovers the standard T-duality transformation. Not less importantly, it also acts globally on the torus bundle with a monodromy description of the supermembrane theory compactified on $M_9\times T^2$ target-space \cite{sculpting, gmmpr, gmpr2}. Previous search of the Supermembrane T-duality generalization were investigated in \cite{dl, russo, aldabe2}.

In this paper we are going to determine explicitly the construction of T-duality transformation, showing as a new result that it may be reduced to a nonlinear realization of a parabolic equivalence subgroup of $SL(2,Z)$ with a dependence on the Kaluza-Klein and winding modes of the theory. This fact will be of relevance when trying to obtain a deeper understanding of the role of T-dualities and gauged supergravities in nine dimensions.
The paper is organized as follows: in section 2 we recall local  T-duality transformation for the supermembrane and its String limit. Since in this note we are only interested in the T-duality construction, we will not discuss global aspects of T-duality. In section 3 the new result is introduced through the construction of T-duality transformation. Section 4 is devoted for conclusions.
%
%
%
%%%%%%%%%%%%%%%%%%%%%%%%%%%%%%%%%%%%%%%%%%%%%%%%%%%%%%%%%%%%%%%%%%%%%%%%%%%%%%%%%%%%%%%%%
%
%
%%%%%%%%%%%%%%%%%%%%%%%%%%%%%%%%%%%%%%%%%%%%%%%%%%%%%%%%%%%%%%%%%%%%%%%%%%%%
\section{T-duality}
In this section we will consider the local and global action of a generalized T-duality at the M-theory level. In particular, at the level of the supermembrane theory it acts globally in a relevant way. In fact, when the supermembrane is compactified on $M_9\times T^2$, the T-duality acts on the moduli of the 2-torus and interchanges Kaluza-Klein charges of the mass operator to the winding ones $W$.

The T-duality transformation \cite{sl2z} is:
\begin{equation}
\label{tt1}
\textrm{for the moduli}: \quad \widetilde{\tau}=\frac{\delta \tau+\beta}{\gamma\tau +\delta}; \quad \mathcal{Z}\widetilde{\mathcal{Z}}=1; \quad
\textrm{for the charges}: \quad \widetilde{\mathcal{Q}}=\Lambda_0 \mathcal{Q},\quad \widetilde{W}=\Lambda_0^{-1}W \> ,
\end{equation}
where we have defined: $\mathcal{Z}:= TA\widetilde{Y}$ and $\mathcal{\widetilde{Z}}:= T \widetilde{A}Y$, with $T$ denoting the supermembrane tension; $\tau$ is a complex moduli $\tau={\rm Re}(\tau) +i{\rm Im}(\tau)$, ${\rm Im}(\tau)>0$ ($\tau$ is the complex coordinate of the Teichm\"uller space for $g=1$); $A = (2\pi R)^2 {\rm Im}(\tau)$ the area of the target torus and $Y=\frac{R \, {\rm Im}(\tau)}{\vert q\tau -p\vert}$. We also have $\Lambda_0=\left( \begin{array}{cc} \delta & \beta \\ \gamma & \delta \end{array}\right) \in SL(2,{Z})$; $\mathcal{Q}=\left(\begin{array}{cc} p \\ q \end{array}\right)$ the Kaluza-Klein charges, and $W=\left(\begin{array}{cc} l_1 & l_2\\ m_1 & m_2\end{array}\right)$ the nontrivial windings of the supermembrane along the 2-torus of the targetspace. In all these expressions the tilde variables are the transformed quantities under T-duality.
T-duality transformations act nonlinearly on the charges of the supermembrane, in distinction with the usual $SL(2,Z)$ transformation. This fact is very relevant for understanding the dual multiplet structure. The KK modes are mapped onto the winding modes and viceversa nonlinearly.

The dual radius of the torus moduli is defined as,
\begin{equation}
\label{trans}
\widetilde{R}= \frac{\vert \gamma\tau+\delta\vert \vert q\tau-p\vert ^{2/3}}{T^{2/3}({\rm Im}(\tau))^{4/3}(2\pi)^{4/3}R} \> .
\end{equation}
In the case $\mathcal{Z}=\mathcal{\widetilde{Z}}=1$ the T-duality transformation becomes a symmetry and it imposes a relation between the tension, the moduli and the KK charges given by:
\begin{equation}
\label{la}
T_0=\frac{\vert q\tau-p\vert}{R^3 ({\rm Im}(\tau))^2} \> .
\end{equation}
Then, if we know the values of the moduli, it fixes the allowed tension $T_0$. Conversely, for a fixed tension $T_0$, the radius, the Teichm\"uller parameter of the 2-torus, and the KK charges satisfy (\ref{la}). Also, for $\mathcal{Z}=1$ the hamiltonian and the mass operator of the supermembrane with central charges is invariant under T-duality:
\begin{equation}
 M^2 = T^2 n^2 A^2 + \frac{m^2}{Y^2}+ T^{2/3}H =\frac{n^2}{\widetilde{Y}^2}+ T^2 m^2 \widetilde{A}^2+ T^{2/3}\widetilde{H} \> ,
\end{equation}
with $H=\widetilde{H}$.

%
%
%%%%%%%%%%%%%%%%%%%%%%%%%%%%%%%%%%%%%%%%%%%%%%%%%%%%%%%%%%%%%%%%%%%%%%%%%
\paragraph{String T-duality transformation limit.}
It was possible to consider the string-like configurations from the physical configurations of the supermembrane with central charges \cite{sl2z,gmpr2,tduality}. In this condition the fields on the torus $\Sigma$, a compact Riemann surface, can be written as:
\begin{equation}
X^{m}=X^{m}(\tau, c_{1}\widehat{X}^{1}+c_{2}\widehat{X}^{2}),\quad  A^{r}=A^{r}(\tau,
c_{1}\widehat{X}^{1}+c_{2}\widehat{X}^{2}) \> ,
\end{equation}
where $c_{1},c_{2}$ are relative prime integral numbers because the global periodicity condition. $X^{m}, A^{r}$ are scalar fields and they may always be expanded on a Fourier basis in term of a double periodic variable of that form. On these configurations we have
\begin{equation}
\{X^{m},X^{n}\}=\{X^{m}, A^{r}\}=\{A^{r},A^{s}\}=0 \> .
\end{equation}
We then obtain the final expression for the mass contribution of the string states \cite{sl2z}:
\begin{equation}
\label{78}
M_{11}^{2}\vert_{SC}= (n T_{11}A)^2+(\frac{m}{Y})^{2}+ 8\pi^{2} R_{11}T_{11}\vert q\tau -p \vert
(N_{L}+N_{R}) \> ,
\end{equation}
\noindent where $(p,q)$ are relatively prime.

The harmonic sector changes \cite{sl2z}, that is:
\begin{equation}
dX_{h}=(qmd\widetilde{X}^{1}+pd\widetilde{X}^{2})+\widetilde{\tau}(-Q n d\widetilde{X}^{1}+Pd\widetilde{X}^{2}) \> ,
\end{equation}
where $p,q$ and $Q,P$ are the winding numbers of the supermembrane. This transformation leaves the hamiltonian invariant. Then, by restricting the worldvolume configurations of the M2 to those of the string \cite{sl2z}, we exactly recover the mass operator of the IIB theory as was formerly introduced by Schwarz \cite{schwarz}, but it has a plus: in the supermembrane with central charges the pure membrane excitations are known. If now a T-duality is performed on the supermembrane M2 mass operator restricted to string-like configurations, then an SL(2,Z) non-perturbative multiplet of IIA is obtained \cite{sl2z}.

Next we recover the standard T-duality transformations for the closed string operator as a limit of the T-duality for the Supermembrane. We define:
\begin{equation}
R_1=\frac{R_{11}^{1/2}}{T_{11}^{1/6}}; \quad R_{2}=R_{11}^{3/2}T_{11}^{1/6} {\rm Im}(\tau) \> ,
\end{equation}
for the torus degenerating into a circle $S^1$.
Since $R_{11}\to 0$, it implies $R_1\to 0$ but $R_2$ is finite, so it corresponds to a closed curve that topologically is a circle. If we re-express the winding condition in terms of the new variables, we have:
\begin{equation}
\oint_{\mathcal{C}_s} dX^1= 2\pi R_1 l_s \> ; \quad \oint_{\mathcal{C}_s} dX^2= 2\pi R_2 m_s \> .
\end{equation}
Because $R_1\to 0$, although $l_s$ is taking finite, the first winding condition vanishes and the only residual winding condition is associated to the $S^1$ modulus is $R_2$. The former T-duality relations of the moduli in this limit become reduced to:
\begin{equation}
  Z\widetilde{Z}=1 \vert_{string} \, , \quad \Rightarrow\quad
   T_{M2}^{4/3} R_2^3\widetilde{R}_2^3=1\to \widetilde{R}_2=\frac{\alpha^{'}}{R_2} \> ,
\end{equation}
where $\widetilde{R}_2= T_{11}^{1/2}\widetilde{A}^{1/2}\widetilde{Y}^{1/2}$. This defines for $T_{(p,q)}=T$ on the IIB string side precisely the duality relation of the strings. The transformation on the charges and windings are given by (\ref{tt1}) and we finally obtain:
\begin{equation}
\{R;(l_1,m)\}\stackrel{T-duality}{\longrightarrow}\{\widetilde{R}=\frac{\alpha^{'}}{R}; (m,l_1)\} \> ,
\end{equation}
where $m$ is the common factor between the Kaluza-Klein charges $(p,q)$.
%
%
%%%%%%%%%%%%%%%%%%%%%%%%%%%%%%%%%%%%%%%%%%%%%%%%%%%%%%%%%%%%%%%%%%%%%%%%%%%%%%%%%%%%%%%%%%%%%%%%%%%%%%%%%%%%%%%%%%%
\section{ Explicit construction of the $\Lambda_0$ transformation}
In this section we are going to deep on the characterization of  the T-duality the map $\Lambda_0$ of the supermembrane. In \cite{gmpr2}, it was shown the existence of the $\Lambda_0$ transformation of the form
\begin{equation}
\Lambda_0=\left(\begin{array}{cc} \delta & \beta\\ \gamma & \delta\end{array}\right) \> \in SL(2,Z),
\end{equation}
with the coefficients $(\delta,\beta,\gamma)$ having a nontrivial dependence on the windings $(l,m)$ and Kaluza Klein charges $(p,q)$, such that
\begin{equation}
\left(\begin{array}{c} l \\ m\end{array}\right) \, =\Lambda_0 \left(\begin{array}{c} p \\ q\end{array}\right) \> .
\end{equation}
Now we are going to show the existence of a parabolic class of solutions. Given the winding matrix $W$, its determinant $det W=n$ corresponds to the central charge of the supersymmetric algebra of the theory. We will assume in the following that $n$ is an integer prime number. The equivalence class of the windings is defined by $\{WS\}$ where $S\in SL(2,Z)$.  Given $\mathcal{Q}=\left(\begin{array}{c} p \\ q\end{array}\right) \, ,$ the Kaluza Klein charges where $(p,q)$ are relatively primes, then there always exists a parabolic transformation,
\begin{equation}
\Lambda_0=\left(\begin{array}{cc} 1 & k\\ 0 & 1\end{array}\right) \> ,
\end{equation}
such that $W=\Lambda_0 Q$, where $Q=\left(\begin{array}{cc} p & P\\ q & Q\end{array}\right) \, $ or $Q=\left(\begin{array}{cc} q & U\\ q & V\end{array}\right) \, $, where $P,Q,U,V$ are integers such that:
\begin{equation}
det \> W= n = det \>  Q \> .
\end{equation}
It is straightforward to show that given $(p,q)$ relatively primes, there always exists $P,Q,U,V$ satisfying $\det Q=n$.
\subsection{Proof}
Given $W$ there always exists $S_1\in SL(2,Z)$ \cite{sl2z}, such that
\begin{equation}
WS_1=\left(\begin{array}{cc} \lambda_1 & \rho\\ 0 & \lambda_2\end{array}\right) \> ,
\end{equation}
with $\rho,\lambda_1,\lambda_2$ integers and satisfying that $\lambda_1\lambda_2=n.$ Let us consider $WS_1S_2$, where
$S_2=\left(\begin{array}{cc} 1 & r\\ 0 & 1\end{array}\right) \,\in SL(2,Z)$, with $r$ an integer to be determined. Then,
\begin{equation}
WS_1S_2=\left(\begin{array}{cc} \lambda_1 & \lambda_1r+\rho\\ 0 & \lambda_2\end{array}\right) \> .
\end{equation}
Finally consider
\begin{equation}
R=\left(\begin{array}{cc} \lambda_1 & s\\ 0 & 1\end{array}\right) \,\in SL(2,Z) \> ,
\end{equation}
then
\begin{equation}
RWS_1S_2=\left(\begin{array}{cc} \lambda_1 & \lambda_1 r+\lambda_2 s+\rho\\ 0 & \lambda_2\end{array}\right) \> .
\end{equation}
Since $n$ is prime, $\lambda_1$ and $\lambda_2$ are relatively primes and one of them is $1$. There always exists $r,s$ such that
\begin{equation}
\lambda_1 r+\lambda_2 s=-\rho \>,
\end{equation}
hence
\begin{equation}
RWS_1S_2=\left(\begin{array}{cc} \lambda_1 &0\\ 0 & \lambda_2\end{array}\right) \> .
\end{equation}
The same argument may be performed with the matrix $Q$:
\begin{equation}
\widetilde{R}Q\widetilde{S}_1\widetilde{S}_2=\left(\begin{array}{cc} \widetilde{\lambda}_1 &0\\ 0 & \widetilde{\lambda}_2\end{array}\right) \> .
\end{equation}
Then either $\lambda_i=\widetilde{\lambda_i}$ or $\lambda_i=\widetilde{\lambda_j}$ with $i\ne j$ for $i,j=1,2.$ In the first case $\lambda_i=\widetilde{\lambda_i}$
\begin{equation}
RWS_1S_2=\widetilde{R}Q\widetilde{S}_1\widetilde{S}_2 \> ,
\end{equation}
which imply
\begin{equation}
WS_1S_2\widetilde{S}_2^{-1}\widetilde{S}_1^{-1}=R^{-1}\widetilde{R}Q \> ,
\end{equation}
with $Q=\left(\begin{array}{cc} p & P\\ q & Q\end{array}\right) \,$  $S_1S_2\widetilde{S}_2^{-1}\widetilde{S}_1^{-1}\in Sl(2,Z)$, and $R^{-1}\widetilde{R}\in SL(2,Z).$\newline
Moreover,
\begin{equation}
\Lambda_0= R^{-1}\widetilde{R}=\left(\begin{array}{cc} 1 & \widetilde{s}-s\\ 0 & 1\end{array}\right) \> ,
\end{equation}
with $k=\widetilde{s}-s$ as claimed.\\If $\lambda_i=\widetilde{\lambda_j}, i\ne j$ the same argument follows by considering
$Q=\left(\begin{array}{cc} q & U\\ q & V\end{array}\right) \, .$\newline

Consequently it always exists a parabolic solution for the T-duality transformation $\Lambda_0$ with a nonlinear dependence on the windings and Kaluza-Klein charges of the theory.

%
%
%%%%%%%%%%%%%%%%%%%%%%%%%%%%%%%%%%%%%%%%%%%%%%%%%%%%%%%%%%%%%%%%%%%%%%%%%%%%%%%%%%%%%%%%%%%%%%%%%%%%%%%%%%%%%%%%%%%%%%%%%%%%%%%%%%%%%%%%%%%%%%%%%%%%%%%
\section{Conclusions}
 In this work we deep on the T-duality characterization of the compactified Supermembrane theory on $M_9\times T^2$. The T-duality transformation realizes non-linearly the SL(2,Z) with a nontrivial dependence on the windings and Kaluza-Klein charges of the theory. Moreover, the novelty of this work, shows that indeed the $\Lambda_0$ transformation always has a solution that belongs to the equivalence class of parabolic subgroup of $SL(2,Z)$. This is a nontrivial result, since the same statement does not necessarily hold for other inequivalent classes of T-duality transformations. This is an intermediate step towards a deeper understanding in terms of a fundamental invariants of M-theory, of the differences in the multiplet structure of the nine dimensional type IIA and type IIB gauged supergravities. The existence of this class of solutions is going to have very relevant implications toward the understanding of the structure of duals in type II gauged supergravities in nine dimensions as it will be discussed further in \cite{gmpr4}.
%
%
%%%%%%%%%%%%%%%%%%%%%%%%%%%%%%%%%%%%%%%%%%%%%%%%%%%%%%%%%%%%%%%%%%%%%%%%%%%%%%%%%%%%%%%%%%%%%%%%%%%%%%%%

\end{document}